\documentclass[12pt]{iopart}
\usepackage{iopams}  
\usepackage{graphicx}
\begin{document}
\pdfminorversion=4
\title[Evolution properties of coherent Gaussian vortex beams through weak oceanic turbulence ]{Evolution properties of coherent Gaussian vortex beams propagating through weak oceanic turbulence and its application for measurement} 

\author{Hantao Wang$^1$, Huajun Zhang$^1$, Mingyuan Ren$^1$,  Jiandong Cai$^1$ and Yu Zhang$^{1,*}$}

\address{School of Physics, Harbin Institute of Technology, No.92, Xidazhi Road, Harbin 150001, China}
\ead{zhangyuhitphy@163.com}
\vspace{10pt}

\begin{abstract}
The cross-spectral density of coherent Gaussian vortex beams propagating through weak oceanic turbulence is derived from extended Huygens-Fresnel principle and Nikishov spectrum. The evolution of a coherent superposition field composed of dual Gaussian vortex beams with $+1$ and $-1$ topological charges respectively through weak oceanic turbulence is investigated in z plane and y plane. It is shown that the non-zero separation distance of two beams in x direction enhances the oceanic turbulence effect on interference light field. The variation of intensity distribution in z plane and on the central axis of two beams in y plane are both related to the strength of oceanic turbulence, separation distance, propagation distance and waist width. The extra fluctuation of the intensity on the central axis of two beams in y plane leads to high sensitivity of oceanic turbulence. This characteristic has potential application in non-contact optical tomography of oceanic turbulence strength along the beam propagation path by lateral scattering intensity.
\end{abstract}


%
%
%
%
%

\section{Introduction}
The effects of oceanic turbulence on optical beams propagation like beam wander \cite{Lu2016Beam}, beam spreading \cite{Wu2016Spreading} and scintillation \cite{Korotkova2012Light} have been studied for many years. In order to achieve the goal of reducing the effect on beam propagation, partially coherent beams have been studied widely \cite{Dogariu2003Propagation,Gbur2002Spreading}. In recent years, as a kind of structured light, beams with phase singularities propagating through oceanic turbulence have been investigated a lot \cite{Cheng2016Propagation,Huang2014Evolution}. Especially in the area of underwater optical communication, partially coherent vortex beams have been paid much attention to for its oceanic turbulence resistance \cite{Li:19}. In addition, a lot of noteworthy properties give them an advantage over Gaussian beams \cite{Soifer2016Vortex}.

However, the vortex beam propagation has been verified to measure turbulence strength by radius of a ring dislocation \cite{Gu2010Measurement}. It inspires a new thought of measuring the strength of oceanic turbulence. Thinking in the reverse direction, drawing on the methods to reduce the influence of oceanic turbulence, coherent vortex beams is suitable to enhance the effect of oceanic turbulence. Besides that, the variance of the wandering decreases with the topological charge of the vortex beam increasing \cite{Aksenov2013The}. And the first-order vortices are more stable than higher-order vortices \cite{Gbur2008Vortex}. These characters lead the first-order vortex beam to be the optimal vortex beam for measuring the oceanic turbulence strength.

Therefore, in this paper, the coherent superposition field composed of the first-order Gaussian vortex beams through oceanic turbulence is investigated. The expressions of the cross-spectral density through oceanic turbulence are derived in \sref{sec2}. Based on that, as the simplest condition, the evolution behavior of average intensity of dual coherent Gaussian vortex beams with $+1$ and $-1$ topological charges respectively in z plane is studied in \sref{sec3}. Then \sref{sec4} illustrates the evolution behavior of average intensity on central axis of dual coherent Gaussian vortex beams in y plane. In the end, the conclusion is shown in \sref{sec5}.

\section{The cross-spectral density of a coherent superposition field composed of Gaussian vortex beams through oceanic turbulence}\label{sec2}

The field distribution of a vortex beam, whose topological charge is $+1$ or $-1$, at the $z = 0$ source plane can be expressed as

\begin{equation}
U_{ \pm }\left( {{{\boldsymbol{\rho'}}} , z = 0} \right) = {E_{ \pm }}\exp \left( { - \frac{{\boldsymbol{\rho'}^{2}}}{{{\sigma ^2}}}} \right)\left( {{{\rho'_x}} {\rm{ \pm }} i{{\rho'_y}}} \right)\exp \left({ i\phi }\right), \label{eq1}
\end{equation}

where $E_{ \pm }$ is the electric field amplitude of the vortex beam,  $\sigma$ is the waist width of the Gaussian background beam, ${{\boldsymbol\rho'}}\equiv\left(\rho'_x,\rho'_y\right)$ is two-dimensional coordinate vectors at source plane and $\phi$ is the initial phase. For coherent superposition field in this paper, the effect on the interference field by the spatial correlation length is not significant for coherent Gaussian vortex beams. Therefore,  the spatial correlation length term is negligible in \eref{eq1}.  

The cross-spectral density of the linear combination of coherent Gaussian vortex beams with $+1$ and $-1$ topological charges at source plane is written as 
\begin{eqnarray}
 {W}\left( {{{\boldsymbol\rho'_1}},{{{\boldsymbol\rho'_2}}},z=0}\right)   =& \Bigg\langle \left[ \sum\limits_{n = 1}^N {{U_{+}}\left( {{{\boldsymbol\rho'_{1n}}},0} \right)}
+ \sum\limits_{m = 1}^M {{U_{-}}\left( {{{\boldsymbol\rho'_{1m}}} ,0} \right)} \right]^{*} \nonumber\\
&\times \left[ \sum\limits_{n = 1}^N {{U_{+}}\left( {{{\boldsymbol\rho'_{2n}}} ,0} \right)}
+ \sum\limits_{m = 1}^M {{U_{-}}\left( {{{\boldsymbol\rho'_{2m}}} ,0} \right)} \right] \Bigg\rangle,
\label{eq2}
\end{eqnarray}
where $N$ is the number of Gaussian vortex beams with $+1$ topological charges, $M$ is the number of Gaussian vortex beams with $-1$ topological charges. ${{\boldsymbol{\rho}}^{'}_{1}}$ and ${{\boldsymbol{\rho}}^{'}_{2}}$ are the positions of two points at source plane respectively. The asterisk * specifies the complex conjugate, $\langle{\cdot}\rangle$ represents the ensemble average. 

Expanding \eref{eq2}, the cross-spectral density can always be derived as the linear combination of four parts
\begin{equation}
\fl {W_{++}}\left( {{{\boldsymbol\rho'_{1}}} ,{\boldsymbol\rho'_{2}},0} \right) = E_ + ^ * {E_ + }\exp \left( { - \frac{{\boldsymbol\rho'^{2}_{1}}{\rm{ + }}{\boldsymbol\rho'^{2}_{2}}}{{{\sigma ^2}}}} \right)\left( {{\rho'_{1x}} - i{{\rho'_{1y}}}} \right)\left( {{{\rho'_{2x}}} + i{\rho'_{2y}}} \right)e^{ \left( i\phi_{2n}-i\phi_{1n}\right)},
\label{eq3}
\end{equation}

\begin{eqnarray}
\fl  {W_{--}}\left( {{{\boldsymbol\rho'_{1}}} ,{\boldsymbol\rho'_{2}},0} \right) =&  {E}_{-}^{*}{E}_{-}\exp \left( { - \frac{{\boldsymbol\rho'^{2}_{1}}{\rm{ + }}{\boldsymbol\rho'^{2}_{2}}}{{{\sigma ^2}}}} \right)\left({\rho'_{1x}} + i{\rho'_{1y}} \right)\left({\rho'_{2x}} - i{\rho'_{2y}}  \right)e^{ \left( i\phi_{2m}-i\phi_{1m}\right)},
\label{eq4}
\end{eqnarray}

\begin{eqnarray}
\fl  {W_{+-}}\left( {{{\boldsymbol\rho'_{1}}} ,{\boldsymbol\rho'_{2}},0} \right) =&  {E}_{+}^{*}{E}_{-}\exp \left( { - \frac{{\boldsymbol\rho'^{2}_{1}}{\rm{ + }}{\boldsymbol\rho'^{2}_{2}}}{{{\sigma ^2}}}} \right)\left( {\rho'_{1x}} - i{\rho'_{1y}} \right)\left( {\rho'_{2x}} - i{\rho'_{2y}}  \right)e^{ \left( i\phi_{2m}-i\phi_{1n}\right)},
\label{eq5}
\end{eqnarray}

\begin{eqnarray}
\fl {W_{-+}}\left( {{{\boldsymbol\rho'_{1}}} ,{\boldsymbol\rho'_{2}},0} \right) =&  {E}_{-}^{*}{E}_{+}\exp \left( { - \frac{{\boldsymbol\rho'^{2}_{1}}{\rm{ + }}{\boldsymbol\rho'^{2}_{2}}}{{{\sigma ^2}}}} \right) \left({\rho'_{1x}} + i{\rho'_{1y}} \right)\left( {\rho'_{2x}}  + i{\rho'_{2y}}\right)e^{ \left( i\phi_{2n}-i\phi_{1m}\right)},
\label{eq6}
\end{eqnarray}
where the subscript '$+$' and '$-$' correspond to topological charge $+1$ and $-1$ respectively. To pay more attention to the evolution of coherent superposition through oceanic turbulence, considering the simplest condition, the $N$ and $M$ are all assumed to be $1$. Therefore, \eref{eq3} and \eref{eq4} represent the cross-spectral densities of Gaussian vortex beams with $+1$ and $-1$ topological charge respectively. And the rest parts are the interference terms. Then for general conditions, the separation distance between the centers of two beams is considered. For easy derivation and symmetry, the projection of separation distance on y axis is assumed to be zero and that on x axis is $2d$. Therefore, \eref{eq5} and \eref{eq6} transform into
\begin{eqnarray}
\fl {W_{+-}}\left( {{{\boldsymbol\rho'_{1}}} ,{\boldsymbol\rho'_{2}},0} \right) =&  {E}_{+}^{*}{E}_{-}\exp \left[ { - \frac{\left({\rho'_{1x}-d}\right)^{2}{\rm{ + }}{\left(\rho'_{2x}+d\right)}^{2}}{{{\sigma ^2}}}} \right]\exp \left( { - \frac{{\rho'^{2}_{1y}}{\rm{ + }}{\rho'^{2}_{2y}}}{{{\sigma ^2}}}} \right)\nonumber\\ &\times \left[\left({\rho'_{1x}}-d\right) - i{\rho'_{1y}} \right]\left[\left( {\rho'_{2x}}+d \right)  - i{\rho'_{2y}}\right]\exp \left( i\phi_{2}-i\phi_{1}\right),
\end{eqnarray}

\begin{eqnarray}
\fl {W_{-+}}\left( {{{\boldsymbol\rho'_{1}}} ,{\boldsymbol\rho'_{2}},0} \right) =&  {E}_{-}^{*}{E}_{+}\exp \left[ { - \frac{\left({\rho'_{1x}+d}\right)^{2}{\rm{ + }}{\left(\rho'_{2x}-d\right)}^{2}}{{{\sigma ^2}}}} \right]\exp \left( { - \frac{{\rho'^{2}_{1y}}{\rm{ + }}{\rho'^{2}_{2y}}}{{{\sigma ^2}}}} \right)\nonumber\\ &\times \left[\left({\rho'_{1x}}+d\right) + i{\rho'_{1y}} \right]\left[\left( {\rho'_{2x}}-d \right)  + i{\rho'_{2y}}\right]\exp \left( i\phi_{2}-i\phi_{1}\right).
\end{eqnarray}
Due to the separation distance only having impact on the intensity distribution of interference terms, \eref{eq3} and \eref{eq4} can still represent the cross-spectral density of Gaussian vortex beams through changing the $\boldsymbol\rho'$ to which contains $d$. 

Using extended Huygens-Fresnel principle, any part of the cross-spectral density function at z plane of the coherent superposition field of Gaussian vortex beams propagating through oceanic turbulence can be expressed as 
\begin{eqnarray}
\fl {W_{ \pm  \pm }}\left( {{\boldsymbol\rho_{1}},{\boldsymbol\rho_{2}},z} \right) = &\frac{{{k^2}}}{{4{\pi ^2}{z^2}}}\mathrm{\int}\mathrm d{\boldsymbol\rho'_{1}}\mathrm{\int}{W_{ \pm  \pm}}\left( {{{\boldsymbol\rho'_{1}}} ,{\boldsymbol\rho'_{2}},0} \right)\exp \left[ { - ik\frac{{{{\left( {{\boldsymbol\rho_{1}} - {\boldsymbol\rho'_{1}}} \right)}^2} - {{\left( {{\boldsymbol\rho_{2}} - {\boldsymbol\rho'_{2}}} \right)}^2}}}{{2z}}} \right]\nonumber\\
&\times \Big \langle{ \exp \left[ {{\psi ^ * }\left( {{\boldsymbol\rho_{1}},{\boldsymbol\rho'_{1}},z} \right) + \psi \left({{\boldsymbol\rho_{2}},{\boldsymbol\rho'_{2}},z} \right)}\right]} \Big \rangle\mathrm d{\boldsymbol\rho'_{2}},
\label{eq9}
\end{eqnarray}
where ${k}$ is the wave number, ${\boldsymbol{\rho_1}}$ and ${\boldsymbol\rho_{2}}$ are the position of two points at the z plane. ${\psi \left({{\boldsymbol\rho,{\boldsymbol\rho'},z}} \right)}$ is the random part of the complex phase of a spherical wave caused by the oceanic turbulence. The ensemble average of a spherical wave in \eref{eq9} can be expressed as \cite{Gbur2002Spreading}
\begin{eqnarray}
\fl \Big \langle{ \exp \left[ {{\psi ^ * }\left( {{\boldsymbol\rho_{1}},{\boldsymbol\rho'_{1}},z} \right) + \psi \left({{\boldsymbol\rho_{2}},{\boldsymbol\rho'_{2}},z} \right)}\right]} \Big \rangle
 =& \exp  \Big\{- {  {k^2} z T\left( {\eta ,\varepsilon ,{\chi _T},\omega } \right)}\Big[ {{{ \left( {{\boldsymbol{\rho}_1} - {\boldsymbol{\rho}_2}} \right)}^2}}\nonumber\\
 & + \Big( {{\boldsymbol{\rho}_1} - {\boldsymbol{\rho}_2}} \Big)\Big( {{\boldsymbol\rho'_{1}} - {\boldsymbol\rho'_{2}}} \Big)  
 + {{\left( {{\boldsymbol\rho'_{1}} - {\boldsymbol\rho'_{2}}} \right)}^2} \Big]  \Big\},
  \label{eq10}
\end{eqnarray}
where
\begin{eqnarray}
T\left( {\eta ,\varepsilon ,{\chi _T},\omega } \right) = \frac{{{{\rm{\pi }}^2}}}{3}\int_0^\infty  {{\kappa ^3}} {\Phi _{\rm{n}}}\left( \kappa  \right){\rm{d}}\kappa,
\label{eq11}
\end{eqnarray}
is a function which is related to spatial power spectrum of the refractive-index fluctuations of the oceanic turbulence, and $\kappa$ is the spatial wave number. For tractable analysis, the spatial power spectrum of oceanic turbulence model utilized in this paper is Nikishov spectrum \cite{Nikishov2000Spectrum}
\begin{eqnarray}
 {\Phi _{\rm{n}}}\left( {\kappa ,\eta ,\varepsilon ,{\chi _T},\omega } \right) = &0.388 \times {10^{ - 8}}{\varepsilon ^{ - {1 \mathord{\left/
 {\vphantom {1 3}} \right.
 \kern-\nulldelimiterspace} 3}}}{\kappa ^{ - {{11} \mathord{\left/
 {\vphantom {{11} 3}} \right.
 \kern-\nulldelimiterspace} 3}}}{\omega ^{ - 2}}{\chi _T}\left[ {1 + 2.35{{\left( {\kappa \eta } \right)}^{{2 \mathord{\left/
 {\vphantom {2 3}} \right.
 \kern-\nulldelimiterspace} 3}}}} \right]\nonumber\\
 &\times \left( {{\omega ^2}{e^{ - {A_T}\delta }} + {e^{ - {A_S}\delta }} - 2\omega {e^{ - {A_{TS}}\delta }}} \right),
 \label{eq12}
\end{eqnarray}
where
\begin{equation}
\delta  = 8.284{\left( {\kappa \eta } \right)^{{4 \mathord{\left/
 {\vphantom {4 3}} \right.
 \kern-\nulldelimiterspace} 3}}} + 12.978{\left( {\kappa \eta } \right)^2},
 \label{eq13}
 \end{equation}
$\eta$ is the inner scale of turbulence, $\varepsilon$ is the rate of dissipation of turbulent kinetic energy per unit mass of fluid, ${\chi_T}$ is the rate of dissipation of mean-square temperature, and $\omega$ is the relative strength of temperature and salinity fluctuation. Simplifying the \eref{eq11} by substituting \eref{eq12} and \eref{eq13} and ${A_T=1.863\times{10^{-2}}}$, ${A_S=1.9\times{10^{-4}}}$, ${A_{TS}=9.41\times{10^{-3}}}$ \cite{Korotkova2012Light}, we can obtain  
 \begin{equation}
\fl T\left( {\eta ,\varepsilon ,{\chi_T},\omega } \right) = 1.2765 \times {10^{ - 8}}{\omega ^{ - 2}}{\varepsilon ^{ - {1 \mathord{\left/
 {\vphantom {1 3}} \right.
 \kern-\nulldelimiterspace} 3}}}{\eta ^{ - {1 \mathord{\left/
 {\vphantom {1 3}} \right.
 \kern-\nulldelimiterspace} 3}}}{\chi _T}\left( {47.5708 - 17.6701\omega  + 6.78335{\omega ^2}} \right).
\end{equation}

Since \eref{eq10} is restricted to weak fluctuation conditions, the Rytov variance representing the irradiance fluctuations associated with an unbounded plane wave should meet the inequality \cite{Wang2016Broadening,Andrews2005Laser}
 \begin{equation}
 \fl \sigma _R^2 = 3.063 \times {10^{ - 7}}{k^{{7 \mathord{\left/
 {\vphantom {7 6}} \right.
 \kern-\nulldelimiterspace} 6}}}{L^{{{11} \mathord{\left/
 {\vphantom {{11} 6}} \right.
 \kern-\nulldelimiterspace} 6}}}{\varepsilon ^{{{ - 1} \mathord{\left/
 {\vphantom {{ - 1} 3}} \right.
 \kern-\nulldelimiterspace} 3}}}{\chi _T}\left( {0.358 - 0.725{\omega ^{ - 1}} + 0.367{\omega ^{ - 2}}} \right) < 1.
 \end{equation}
For simplicity, the variable parameter representing the effect strength of oceanic turbulence on Gaussian vortex beams is selected to be ${\chi_T}$ due to its linear relations to $T\left( {\eta ,\varepsilon ,{\chi_T},\omega } \right)$ and $\sigma _R^2$. Then other parameters are confirmed to be 
$\eta = 10^{-3} \rm{m}$, $\varepsilon=10^{-7}\rm{m^2{/}s^3}$, and $\omega=-2.5$, which lead ${\chi_T}$ to range from ${10^{-10}\rm{K^2/s}}$ to ${10^{-7}\rm{K^2/s}}$ \cite{thorpe2005turbulent}.

According to the integral formula \cite{Gradshteyn2007Table}
\begin{equation}
\fl \int_{ - \infty }^\infty  {{x^n}\exp \left( { - p{x^2} + 2qx} \right)} {\rm{d}}x = n!\exp \left( {\frac{{{q^2}}}{p}} \right)\sqrt {\frac{{\rm{\pi }}}{p}} {\left( {\frac{q}{p}} \right)^n}\sum\limits_{k = 0}^{\left\lfloor {{n \mathord{\left/
 {\vphantom {n 2}} \right.
 \kern-\nulldelimiterspace} 2}} \right\rfloor } {\frac{1}{{\left( {n - 2k} \right)!\left( k \right)!}}{{\left( {\frac{p}{{4{q^2}}}} \right)}^k}},
\end{equation}
when the initial phase difference is assumed to be zero, \eref{eq9} can be transformed into four equations 
\begin{eqnarray}
\fl {W_{++}}\left( {{{\boldsymbol{\rho}}_1} ,{{\boldsymbol{\rho}}_2},z} \right) =& \frac{{{k^2}E_ + ^2}}{{4\pi {z^2}}}\frac{1}{{p_1^{{3 \mathord{\left/
 {\vphantom {3 2}} \right.
 \kern-\nulldelimiterspace} 2}}p_2^{{3 \mathord{\left/
 {\vphantom {3 2}} \right.
 \kern-\nulldelimiterspace} 2}}}}\exp \left[ { - {k^2}zT{{\left( {{{\boldsymbol{\rho}}_1} - {{\boldsymbol{\rho}}_2}} \right)}^2}} \right]\exp \left( { - ik\frac{{\boldsymbol{\rho}^2_1} - {\boldsymbol{\rho}^2_2}}{2z}} \right)\nonumber\\
  &\times\exp \left( {\frac{{\boldsymbol{q}_1^2}}{{{p_1}}}{\rm{ + }}\frac{{\boldsymbol{q}_2^2}}{{{p_2}}}} \right)\left[ \left({q_{1x}} - i{q_{1y}} \right) \left({q_{2x}} + i{q_{2y}} \right){\rm{ + }}\frac{{{k^2}zT}}{2}\left( {1 + \frac{{\boldsymbol{q}_2^2}}{{{p_2}}}} \right) \right],
  \label{eq17}
\end{eqnarray}

\begin{eqnarray}
\fl {W_{--}}\left( {{{\boldsymbol{\rho}}_1} ,{{\boldsymbol{\rho}}_2},z} \right) =& \frac{{{k^2}E_ - ^2}}{{4\pi {z^2}}}\frac{1}{{p_1^{{3 \mathord{\left/
 {\vphantom {3 2}} \right.
 \kern-\nulldelimiterspace} 2}}p_2^{{3 \mathord{\left/
 {\vphantom {3 2}} \right.
 \kern-\nulldelimiterspace} 2}}}}\exp \left[ { - {k^2}zT{{\left( {{{\boldsymbol{\rho}}_1} - {{\boldsymbol{\rho}}_2}} \right)}^2}} \right]\exp \left( { - ik\frac{{\boldsymbol{\rho}^2_1} - {\boldsymbol{\rho}^2_2}}{2z}} \right)\nonumber\\
  &\times\exp \left( {\frac{{\boldsymbol{q}_1^2}}{{{p_1}}}{\rm{ + }}\frac{{\boldsymbol{q}_2^2}}{{{p_2}}}} \right)\left[ \left({q_{1x}} + i{q_{1y}} \right) \left({q_{2x}} - i{q_{2y}} \right){\rm{ + }}\frac{{{k^2}zT}}{2}\left( {1 + \frac{{\boldsymbol{q}_2^2}}{{{p_2}}}} \right) \right],
  \label{eq18}
\end{eqnarray}

\begin{eqnarray}
\fl {W_{+-}}\left( {{{\boldsymbol{\rho}}_1} ,{{\boldsymbol{\rho}}_2},z} \right) =& \frac{{{k^2}{E}_{+}^{*}{E}_{-}}}{{4\pi {z^2}}}\frac{1}{{p_1^{{3 \mathord{\left/
 {\vphantom {3 2}} \right.
 \kern-\nulldelimiterspace} 2}}p_2^{{3 \mathord{\left/
 {\vphantom {3 2}} \right.
 \kern-\nulldelimiterspace} 2}}}}\exp \left[ { - {k^2}zT{{\left( {{{\boldsymbol{\rho}}_1} - {{\boldsymbol{\rho}}_2}} \right)}^2}} \right]\exp \left( { - ik\frac{{\boldsymbol{\rho}^2_1} - {\boldsymbol{\rho}^2_2}}{2z}} \right)\nonumber\\
  &\times \exp\left({-\frac{2d^2}{\sigma^2}}\right)\exp \left[ {\frac{{{\left({\boldsymbol{q}_{1x}{\rm{+}}D_{1x}}\right)}^2}}{{{p_1}}}{\rm{ + }} \frac{{{\left({\boldsymbol{q}_{2x}{\rm{+}}D_{2x}}\right)}^2}}{{{p_2}}}} \right]
  \exp \left( {\frac{{\boldsymbol{q}_{1y}^2}}{{{p_1}}}{\rm{ + }}\frac{{\boldsymbol{q}_{2y}^2}}{{{p_2}}}} \right)\nonumber\\
  &\times \bigg\{ \bigg. 
  \left[dp_1-dk^2zT{\rm{ + }} \left({\boldsymbol{q}_{1x}{\rm{+}}D_{1x}}\right) - i{q_{1y}} \right]\left[{\left({\boldsymbol{q}_{2x}{\rm{+}}D_{2x}}\right)} - i{q_{2y}} \right]\nonumber\\
  &\left.{\rm{ + }}\frac{{{k^2}zT}}{p_2}\left[ {\left({\boldsymbol{q}_{2x}{\rm{+}}D_{2x}}\right) - i{q_{2y}}} \right]^{2} -dp_2\left[dp_1{\rm{ + }}\left({\boldsymbol{q}_{1x}{\rm{+}}D_{1x}}\right)-iq_{1y}\right] \right\},
  \label{eq19}
\end{eqnarray}

\begin{eqnarray}
\fl {W_{-+}}\left( {{{\boldsymbol{\rho}}_1} ,{{\boldsymbol{\rho}}_2},z} \right) =& \frac{{{k^2}{E}_{-}^{*}{E}_{+}}}{{4\pi {z^2}}}\frac{1}{{p_1^{{3 \mathord{\left/
 {\vphantom {3 2}} \right.
 \kern-\nulldelimiterspace} 2}}p_2^{{3 \mathord{\left/
 {\vphantom {3 2}} \right.
 \kern-\nulldelimiterspace} 2}}}}\exp \left[ { - {k^2}zT{{\left( {{{\boldsymbol{\rho}}_1} - {{\boldsymbol{\rho}}_2}} \right)}^2}} \right]\exp \left( { - ik\frac{{\boldsymbol{\rho}^2_1} - {\boldsymbol{\rho}^2_2}}{2z}} \right)\nonumber\\
  &\times \exp\left({-\frac{2d^2}{\sigma^2}}\right)\exp \left[ {\frac{{{\left({\boldsymbol{q}_{1x}{\rm{-}}D_{1x}}\right)}^2}}{{{p_1}}}{\rm{ + }} \frac{{{\left({\boldsymbol{q}_{2x}{\rm{-}}D_{2x}}\right)}^2}}{{{p_2}}}} \right]
  \exp \left( {\frac{{\boldsymbol{q}_{1y}^2}}{{{p_1}}}{\rm{ + }}\frac{{\boldsymbol{q}_{2y}^2}}{{{p_2}}}} \right)\nonumber\\
  &\times \bigg\{ \left[-dp_1+dk^2zT{\rm{ + }} \left({\boldsymbol{q}_{1x}{\rm{-}}D_{1x}}\right) + i{q_{1y}} \right]\left[{\left({\boldsymbol{q}_{2x}{\rm{-}}D_{2x}}\right)} + i{q_{2y}} \right]\nonumber\\
  &\left.{\rm{ + }}\frac{{{k^2}zT}}{p_2}\left[ {\left({\boldsymbol{q}_{2x}{\rm{-}}D_{2x}}\right) + i{q_{2y}}} \right]^{2} +dp_2\bigg. \left[-dp_1{\rm{ + }}\left({\boldsymbol{q}_{1x}{\rm{-}}D_{1x}}\right)+iq_{1y}\right] \right\},
  \label{eq20}
\end{eqnarray}
where
\begin{equation}
{p_1} = \frac{1}{{{\sigma ^2}}} + {k^2}zT - \frac{{ik}}{{2z}},
\end{equation}

\begin{equation}
{p_2} = \frac{1}{{{\sigma ^2}}} + {k^2}zT + \frac{{ik}}{{2z}} - \frac{{{k^4}{z^2}{T^2}}}{{{p_1}}},
\end{equation}

\begin{equation}
{{\boldsymbol{q}_1}} =  \frac{1}{2}{k^2}zT\left( {{{\boldsymbol{\rho}}_1} - {{\boldsymbol{\rho}}_2}} \right)- \frac{{ik}}{{2z}}{{\boldsymbol{\rho}}_2},
\end{equation}

\begin{equation}
{{\boldsymbol{q}_2}} = - \frac{1}{2}{k^2}zT\left( {{{\boldsymbol{\rho}}_1} - {{\boldsymbol{\rho}}_2}} \right) + \frac{{ik}}{{2z}}{{\boldsymbol{\rho}}_1} + \frac{{{{\boldsymbol{q}_1}}{k^2}zT}}{{{p_1}}},
\end{equation}

\begin{equation}
{{D_{1x}} =  - \frac{d}{{\sigma}^{2}}},
\end{equation}

\begin{equation}
{{D}_{2x}} =  - \frac{{k^2}zTd}{{\sigma}^2p_1} + \frac{d}{{\sigma}^{2}},
\end{equation}
and ${{{\boldsymbol{q_1}}}\equiv\left(q_{1x},q_{1y}\right)}$, ${{{\boldsymbol{q_2}}}\equiv\left(q_{2x},q_{2y}\right)}$.

Equations \eref{eq17}-\eref{eq20} provide the basic analytical solutions to study the evolution of two coherent Gaussian vortex beams with ${\pm{1}}$ topological charges propagating through weak oceanic turbulence. Aiming at the interference parts, the intensity distribution not only is related to the strength of oceanic turbulence and propagation distance $z$ closely, but also is connected with $d$ significantly. What's more, it is clear that the exponential terms in Equations \eref{eq17}-\eref{eq20} denote the Gaussian background beam propagating through oceanic turbulence, and the rest parts represent the effect of phase singularity evolution on beam propagation. 

\section{The evolution of average intensity of a coherent superposition field composed of Gaussian vortex beams through oceanic turbulence}\label{sec3}

Considering the most simplified condition, when ${{\boldsymbol{\rho}_1}}={\boldsymbol{\rho}_2}={\boldsymbol{\rho}}$, $d = 0$ and $E_{+}=E_{-}=E_0$ in Equations \eref{eq17}-\eref{eq20}, the intensity distribution at z plane of the coherent superposition transform into the $x$ component of single Gaussian vortex beam with ${\pm{1}}$ topological charge
\begin{eqnarray}
{I}\left( {{{\boldsymbol{\rho}}},z} \right) =& \frac{{{k^2}E_0 ^2}}{{4\pi {z^2}}}\frac{1}{{p_1^{{3 \mathord{\left/
 {\vphantom {3 2}} \right.
 \kern-\nulldelimiterspace} 2}}p_2^{{3 \mathord{\left/
 {\vphantom {3 2}} \right.
 \kern-\nulldelimiterspace} 2}}}}\exp \left( {\frac{{q_1^2}}{{{p_1}}}{\rm{ + }}\frac{{q_2^2}}{{{p_2}}}} \right)\left[ {q_{1x}} {q_{2x}}{\rm{ + }}\frac{{{k^2}zT}}{2}\left( {1 + \frac{{2q_{2x}^2}}{{{p_2}}}} \right) \right].
\end{eqnarray}
Obviously, under this condition, the intensity distribution has no relation to separation distance which only influences the coherent parts in \eref{eq19} and \eref{eq20}. But when $d$ is non-zero, because of the phase difference fluctuation generated by the oceanic turbulence on different propagation paths of two beams, the interference terms fade away as the strength increase of oceanic turbulence when $d$ and $z$ are fixed. 

\begin{figure}[ht]
\centering\includegraphics[width=0.767\linewidth]{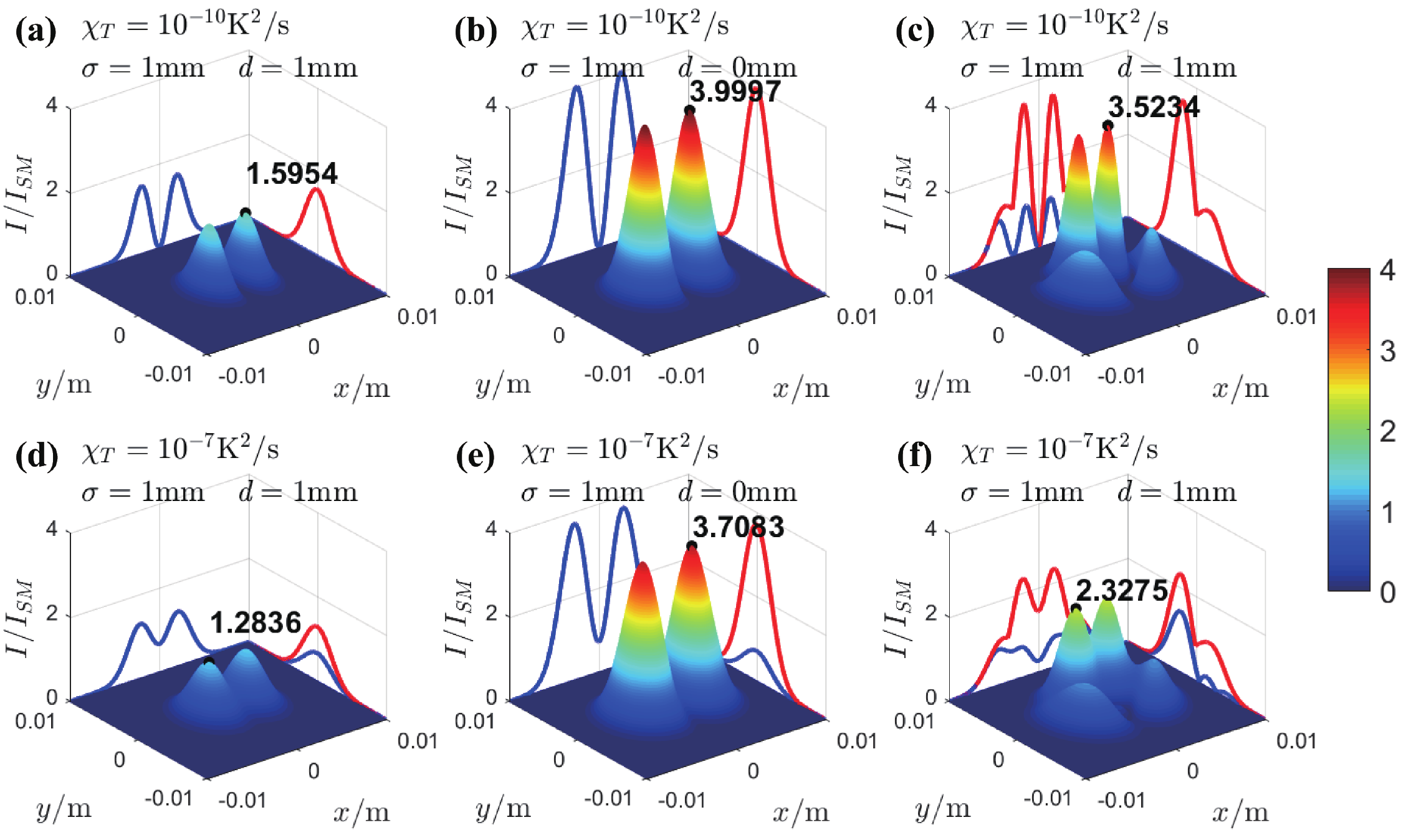}
\caption{The average interference intensity distributions of two coherent Gaussian beams and two coherent Gaussian vortex beams with $+1$ (right) and $-1$ (left) topological charges respectively through oceanic turbulence with different strength.}
\label{figure1}
\end{figure}

To elaborate the evolution of decoherence increase due to $d$, \fref{figure1} shows the average interference intensity distributions of two coherent Gaussian beams and two coherent Gaussian vortex beams ($\sigma = 1mm$) with $+1$ and $-1$ topological charges respectively through oceanic turbulence with different strength. In figures 1(a)-1(c), $ {\chi_T}=10^{-10}\rm{K^2/s}$ and in figures 1(d)-1(f) $ {\chi_T}=10^{-7}\rm{K^2/s}$. In order to present the evolution effect of phase singularity, figure 1(a) and figure 1(d) illustrate the evolution of the average intensity distribution of two coherent Gaussian beams as contrasts, which are the background of the beams in figure 1(c) and figure 1(f). To give prominence to coherent peak, the intensity distribution $I/I_{SM}$ ($I_{SM}$---maximum intensity of single beam) has been normalized based on the maximum intensity of single beam which composes the interference light field. To make it easier to compare, the peak contour projection on x plane and y plane (red lines) and intensity contours at $x=0\rm{mm}$ and $y=0\rm{mm}$ (blue lines) are shown in each sub-graphs. Besides, the peak values of intensity are pointed out in figure 1. 

It is evident that the coherent peak value decreases with the increase of ${\chi_T}$. Compared to non-zero $d$ in figure 1(c) and figure 1(f), the decreasing velocity of peak value in figure 1(b) and figure 1(e) is slower. In addition, through comparing the blue lines, the fusion of two intensity peak is more susceptible when $d$ is non-zero. Therefore, the separation distance leads the light field to be more sensitive to oceanic turbulence. On the surface, the evolution of average intensity distribution represents the decreasing of coherent degree of light caused by oceanic turbulence. Essentially, the decline of intensity peak is mainly attributed to the fluctuation of relative phase difference on the propagation paths. When non-zero $d$ is fixed, the stronger the strength of oceanic turbulence is, the greater the fluctuation of phase difference is. The fluctuation of phase difference leads to the intensity peak value varies in a wide range. That finally results in the decline of average intensity peak value. And this characteristic is shown clearly in figure 1(c) and figure 1(f). 

Then focus on the effect of the evolution of phase singularity, the average intensity peaks in figure 1(c) and figure 1(f) are higher than that in figure 1(a) and figure 1(d), respectively. And the decreasing velocity of peak value is rapider when phase singularity exists. What's more, owing to the existence of phase singularity, the interference intensity distribution is non-centrosymmetric in y direction when $d$ is non-zero. And this character fades away gradually with the strength of oceanic turbulence increasing. These phenomena are mainly determined by the evolution of phase singularity through oceanic turbulence. Non-zero $d$ brings about varying degrees of partial superposition of two beams because of the relative beam wander effect caused by oceanic turbulence. And it is known that the phase distribution around phase singularity is changeable. Therefore, variation of partial superposition generates extra phase difference fluctuation which leads to stronger fluctuation of intensity peak value and then presents the greater decline of average intensity peak. In other words, the extra phase difference fluctuation is only related to the wavefront superposition of two beams. That makes the coherent superposition field composed of Gaussian vortex beams more sensitive to the strength of oceanic turbulence.

Since the average interference intensity peak is not only related to the average electromagnetic intensities of two beams at the same point, but also correlated with the phase difference of two beams, non-zero $d$ and phase singularity result in greater variation of interference intensity peak when beams propagate through oceanic turbulence.  

\begin{figure}[!htbp]
\centering\includegraphics[width=\linewidth]{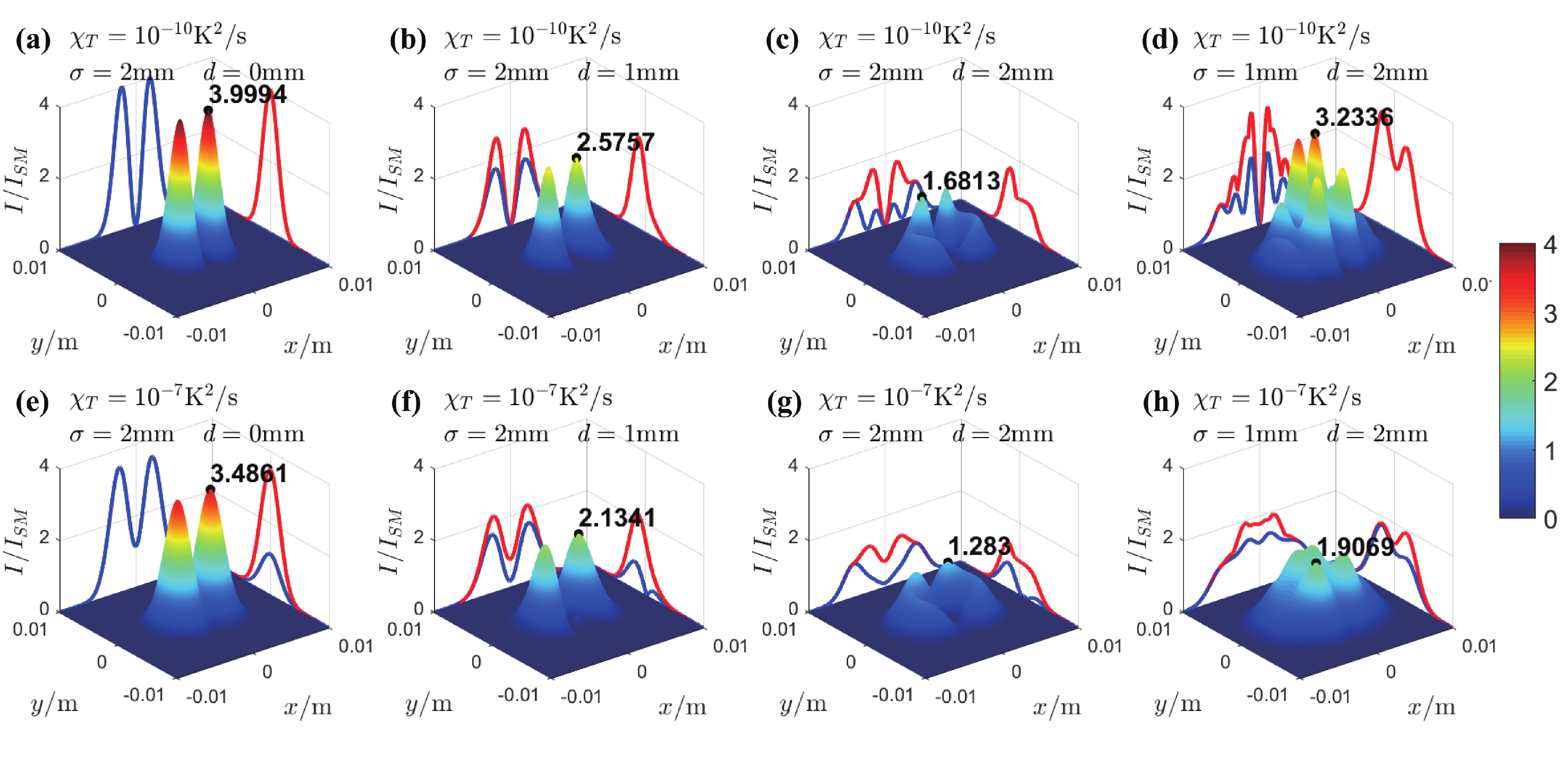}
\caption{The average interference intensity distribution of two coherent Gaussian vortex beams with $+1$ (right) and $-1$ (left) topological charges respectively through oceanic turbulence with different $d$ and $\sigma$.}
\label{figure2}
\end{figure}

Besides the $d$, the waist width $\sigma$ influences the coherent superposition as well. Based on figure 1, we import extra parameter $\sigma$ to investigate the evolution of average interference intensity with different $d$ and different oceanic turbulence strength. Figure 2 illustrates the average interference intensity distribution of two coherent Gaussian vortex beams with $+1$ (right) and $-1$ (left) topological charges respectively through oceanic turbulence with different $d$ and $\sigma$. The effects of oceanic turbulence and $d$ on the evolution of average intensity distribution have already been analyzed in figure 1. Therefore, here only the effect of $\sigma$ is investigated under different conditions. Through comparing the conditions with different $\sigma$ and same $d$ in figure 1 and figure 2, smaller $\sigma$ leads to more distinct diffraction. In general, the diffraction brings about extra overlap of two coherent Gaussian vortex beams and that increases the intensity peak value of interference pattern. In summary, weaker oceanic turbulence, smaller $\sigma$ and smaller $d$ cause greater interference intensity value respectively. However, in figure 2(d) and figure 2(h), the original intensity peak in figure 2(d) decline to the height below the intensity peak in figure 2(h) with the increase of ${\chi_{T}}$. The cause of it is the fusion of two intensity peaks in Fig. 2(h) due to greater beam wander effect. Therefore, the evolution of average interference intensity of two coherent Gaussian vortex beams is influenced by the combination effect of $d$, $\sigma$ and oceanic turbulence.

With the import of non-zero $d$, the evolution of average interference intensity is influenced by oceanic turbulence more significantly. For this characteristic, the coherent superposition field composed of dual Gaussian vortex beams with $+1$ and $-1$ topological charges is more suitable for measuring the strength of oceanic turbulence than single vortex beam. As above mentioned discussion, the parameters of light field and strength of oceanic turbulence jointly determined the evolution of average intensity of coherent superposition field. So the optimum parameters of light field and its standard characteristic of average intensity should be chosen to reflect the variation of oceanic turbulence. 

\begin{figure}[!htbp]
\centering\includegraphics[width=0.8\linewidth]{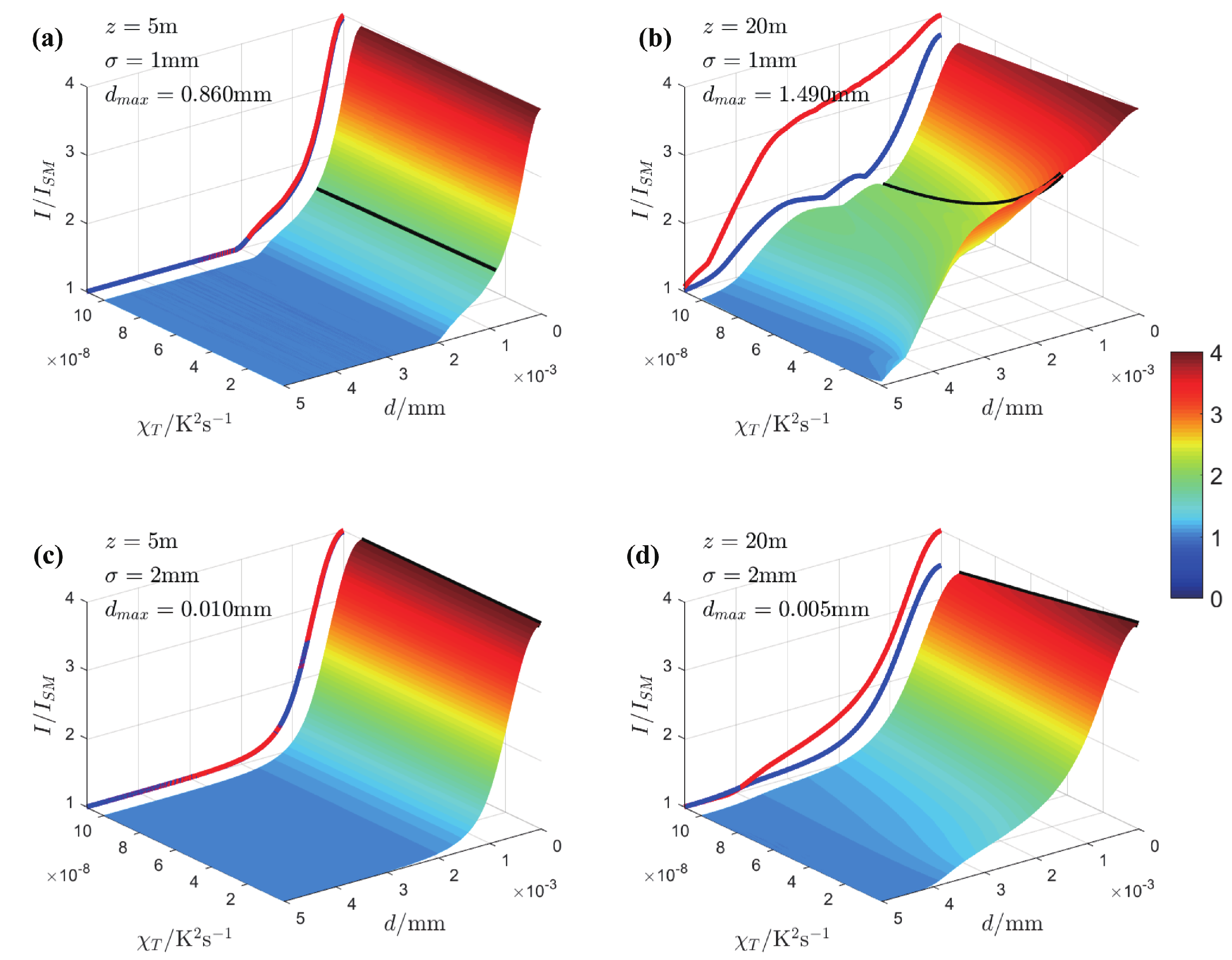}
\caption{The relation of oceanic turbulence strength and $d$ to maximum value of intensity at $z=5\rm{m}$ and $z=20\rm{m}$ with $\sigma=1\rm{mm}$ and $\sigma=2\rm{mm}$ respectively.}
\label{figure3}
\end{figure}

According to above mentioned study, the optimum parameter to denote the influence of oceanic turbulence on the intensity distribution at z plane seems to be the maximum value of intensity. So here presents the relation of oceanic turbulence strength, $\sigma$, $d$ and propagation distance $z$ to maximum value of intensity in figure 3. The red lines denote the changing curve projections of maximum intensity value with $d$ when $\chi_T=10^{-10}{K^2/s}$. Similarly, blue lines are the changing curve projections when $\chi_T=10^{-7}{K^2/s}$. And black lines represent the changing curves of maximum intensity value with oceanic turbulence strength at the difference of blue line and red line reaching the greatest, and $d = d_{max}$. In other words, when $d$ reaches $d_{max}$, the light field has optimum response to oceanic turbulence strength variety at z plane. For this feature, the maximum intensity can be expected to have a potential application in the transmission measurement of oceanic turbulence strength. However, the existence of the fusion of dual intensity peaks in figure 3(b) leads to the non-monotonic relation of maximum intensity value to $d$. This phenomenon restrict the optimum responsiveness to further improve. Besides, for a predetermined light field, the optimum response to oceanic turbulence strength at different $z$ is discrepant. When propagation distance is $5\rm{m}$ in figure 3(a) and figure 3(c), the red lines and blue lines almost overlap together in each sub-graph, respectively. In this case, the average maximum intensity loses the ability to characterize the strength of oceanic turbulence. Therefore, the optimum combination of light field to characterize the strength of oceanic turbulence needs to take into account both $z$ and $\sigma$.

\section{The evolution of average lateral intensity of a coherent superposition field composed of Gaussian vortex beams through oceanic turbulence }\label{sec4}

After investigating the evolution properties of maximum intensity in z plane, we in turn look at the average lateral intensity of interference light field in x plane or y plane.

\begin{figure}[!htbp]
\centering\includegraphics[width=\linewidth]{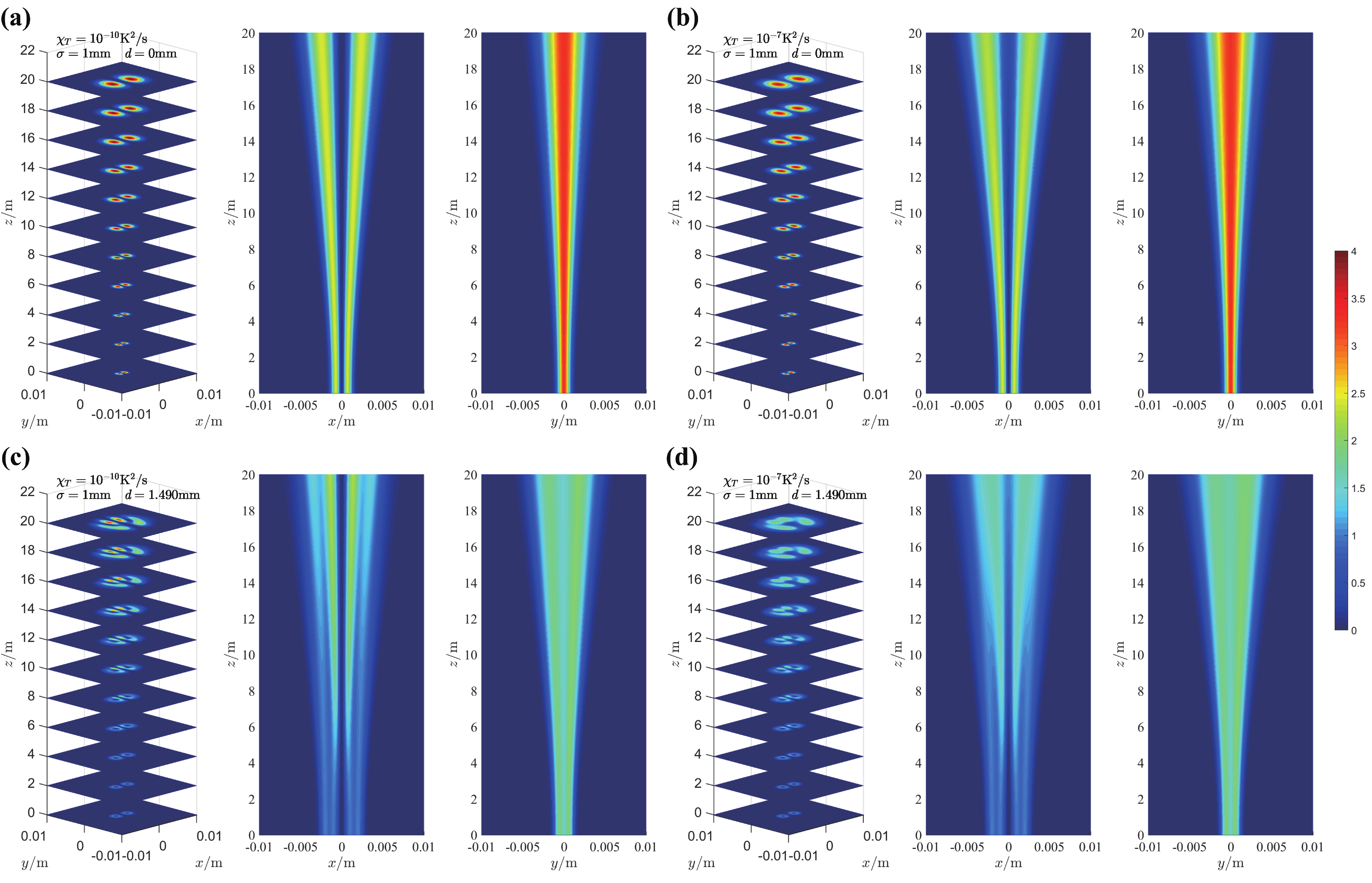}
\caption{The average intensity evolution of dual coherent Gaussian vortex beams and its average lateral intensity distribution evolution through oceanic turbulence.}
\label{figure4}
\end{figure}

To obtain a visible evolution of the average lateral intensity of dual coherent Gaussian vortex beams, we choose the parameters in figure 3(b) where $\sigma=1\rm{mm}$ and $d = 1.490\rm{mm}$ to have a good response to oceanic turbulence strength at $z=20\rm{m}$. The average intensity evolution of dual coherent Gaussian vortex beams and its average lateral intensity distribution evolution through oceanic turbulence are illustrated in figure 4. The image on the left in each sub-graph is the evolution of average intensity distribution in z plane. On the right side of it are two lateral intensity distribution in x plane and y plane, which is obtained by integrating the x and y component of intensity distribution in z plane, respectively. For comparison, the evolution of average intensity distribution under the same conditions, except changing $d$ to be zero, through different strength of oceanic turbulence is present in figures 4(a)-4(b). With the increase of $\chi_{T}$, the intensity distributions in figures 4(c)-4(d) have greater change than these in figures 4(a)-4(b). Especially the variety on the central axis of two beams at y plane enhances more distinct with the propagation distance increase.

The oceanic turbulence perturbs the stable interference field on the central axis of two beams in free space. Because of the accumulated effect of oceanic turbulence on light propagation, the variety of interference light field in the distance is far more greater than it in the vicinity. Besides, it is obviously that the perturbation is distinct in y plane rather than x plane. The reason is $d$ only enhances the effect of oceanic turbulence on the x component of interference intensity distribution. In y plane, due to the lateral intensity on the central axis is minimum, the variety caused by oceanic turbulence is more visible. In short, the lateral intensity is suitable for reflecting the variety of oceanic turbulence strength.

To verify the reliability of lateral intensity in y plane to characterize the strength of oceanic turbulence, the intensity on central axis is analyzed under different strength of oceanic turbulence. The average lateral intensity distribution in y plane can be expressed as
\begin{eqnarray}
 I_{LI} =& \frac{E_0^2{k^2}}{\sqrt {\rm{\pi }}{z^2}}\frac{1}{{p_1^{{3 \mathord{\left/
 {\vphantom {3 2}} \right.
 \kern-\nulldelimiterspace} 2}}p_2^{{3 \mathord{\left/
 {\vphantom {3 2}} \right.
 \kern-\nulldelimiterspace} 2}}}}\frac{1}{{\sqrt {{p_d}} }}\exp \left( { - {p_d}{d^2}} \right)
\left\{ - \left[ {\frac{{ik}}{{4z}} + \frac{{{k^2}zT}}{{8{p_2}}}{{\left( {\frac{k}{z} - \frac{{{k^3}T}}{{{p_1}}}} \right)}^2}} \right]\frac{1}{{4{p_d}}}\right.  \nonumber\\
 & + \left.  \frac{{{k^2}zT}}{2} + {{d^2}\left[ {\left( {\frac{{{k^2}}}{{4{z^2}}} - \frac{{{k^4}T}}{{4z{p_1}}}} \right) - {{\left( {\frac{k}{{2z}} - \frac{{{k^3}T}}{{2{p_1}}}} \right)}^2}\frac{{{k^2}zT}}{{2{p_2}}}} \right]}  \right\} \nonumber\\
 &- \frac{{E_0^2{k^2}}}{{\sqrt {\rm{\pi }} {z^2}}}\frac{1}{{p_1^{{3 \mathord{\left/
 {\vphantom {3 2}} \right. \kern-\nulldelimiterspace} 2}}p_2^{{3 \mathord{\left/
 {\vphantom {3 2}} \right. \kern-\nulldelimiterspace} 2}}}}\frac{1}{{\sqrt {{p_d}} }}\exp \left( { - \frac{{2{d^2}}}{{{\sigma ^2}}}} \right)\exp \left\{ {\frac{{{d^2}  }}{{{\sigma ^4}}}\left[ {\frac{{1  }}{{{p_1}}} + \frac{{{{\left( {1 - \frac{{{k^2}zT}}{{{p_1}}}} \right)}^2}}}{{{p_2}}}} \right]} \right\} \nonumber\\
 &\times  \left\{ {d^2}\left[ {\frac{{{k^2}zT}}{{{p_2}}}\left( {\frac{1}{{{\sigma ^2}}} - \frac{{{k^2}zT}}{{{\sigma ^2}{p_1}}}} \right) + {p_1} - \frac{1}{{{\sigma ^2}}}} \right]\left( {{p_2} - \frac{1}{{{\sigma ^2}}}+ \frac{{{k^2}zT}}{{{\sigma ^2}{p_1}}}} \right)\right. \nonumber\\
 &\left.- \left[ {\frac{{ik}}{{4z}} + \frac{{{k^2}zT}}{{4{p_2}}}{{\left( {\frac{k}{z} - \frac{{{k^3}T}}{{{p_1}}}} \right)}^2}} \right]\frac{1}{{4{p_d}}} \right\} ,\label{eq28}
\end{eqnarray}
where
\begin{eqnarray}
{p_d} = \frac{{{k^2}}}{{4{z^2}}}\left[ {\frac{1}{{{p_1}}} + \frac{{{{\left( {1 - \frac{{{k^2}zT}}{{{p_1}}}} \right)}^2}}}{{{p_2}}}} \right].
\end{eqnarray}
It can be seen that the average lateral intensity is composed of an incoherent part (the first two lines in \eref{eq28}) and a coherent part (the last three lines in \eref{eq28}). And it is quite obvious that when $T = 0$ and $d = 0$, $I_{LI}$ is identically equal to zero. However, when $T$ and $d$ are non-zero, then the coherent part fades away gradually with $z$ increasing. In another word, through analyzing the $I_LI$, the strength of oceanic turbulence can be measured when $d$ is fixed.

\begin{figure}[!htbp]
\centering\includegraphics[width=0.8\linewidth]{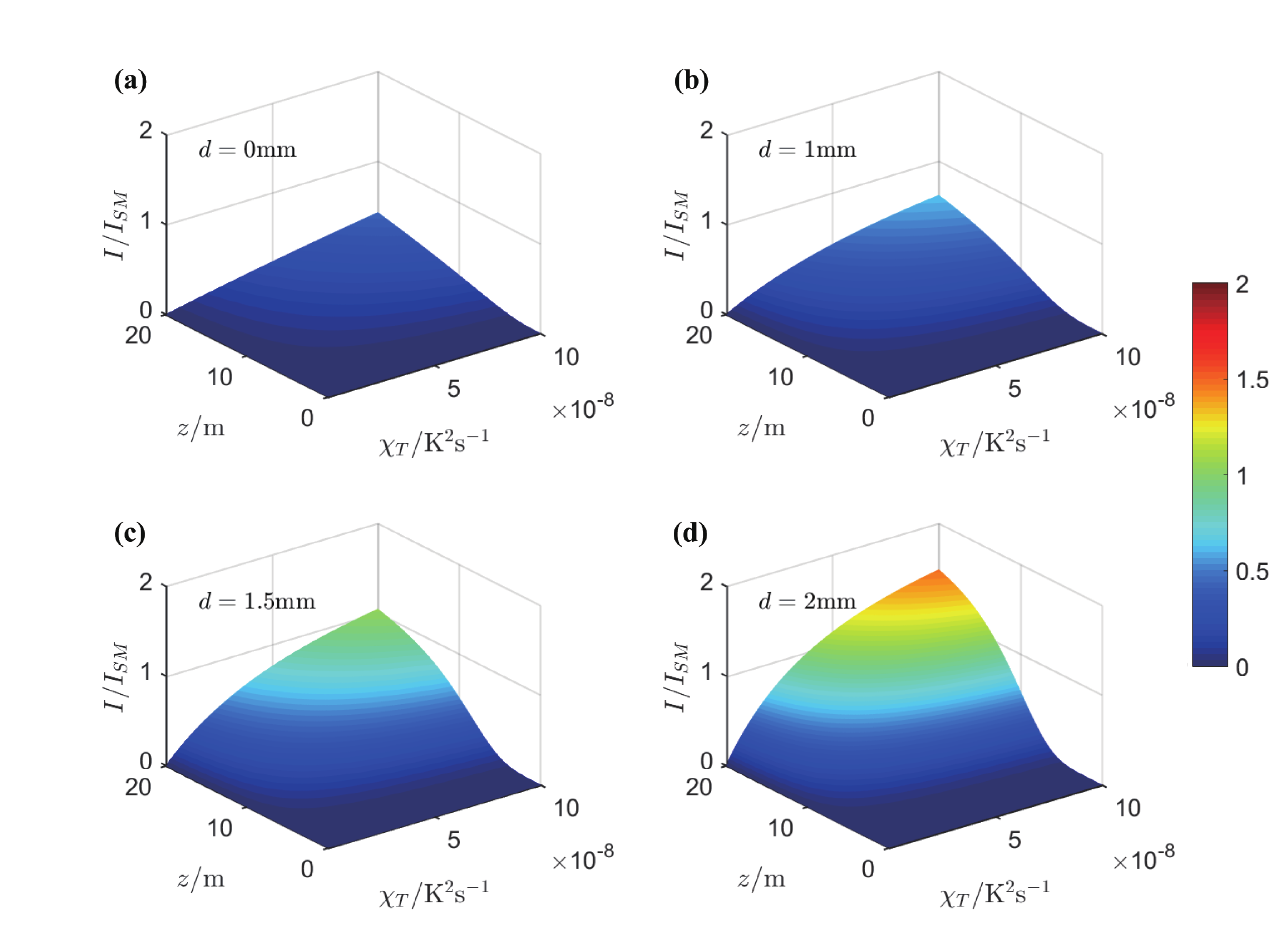}
\caption{The average lateral intensity evolution of dual coherent Gaussian vortex beams with $+1$ and $-1$ topological charges on central axis in y plane through oceanic turbulence.}
\label{figure5}
\end{figure}

To make it more intuitive, the average lateral intensity evolution of dual coherent Gaussian vortex beams with $+1$ and $-1$ topological charges on central axis in y plane through oceanic turbulence is illustrated in figure 5. Comparing to figure 5(a), non-zero $d$ enhances the effect of oceanic turbulence on coherent superposition in figures 5(b)-5(d). In each sub-graph, the relations of intensity value on central axis to oceanic turbulence and propagation distance $z$ are both monotonic. This feature is appropriate for characterizing the strength of oceanic turbulence when the form of beams and propagation distance are determined. Therefore, it has a potential application in the lateral scattering measurement of oceanic turbulence strength. Not only that, the intensity is also increasing by almost two times of the maximum lateral intensity of single beam in figure 5(d). In other words, the average intensity distribution gradually approaches to incoherent superposition. This intensity increase makes perceiving the strength of oceanic turbulence through lateral much more easier. However, the limitation of it could not be ignored as well. In figures 5(b)-5(d), with the increasing of $d$, the range of slow increase of intensity from the transmitter side extends gradually because of barely superposition part of two beams. With Rytov variance increasing, the light field approaches to a superposition of two Gaussian beams. The influence of interference part fades away which leads to scarcely any change of intensity on central axis through oceanic turbulence.

Therefore, the lateral intensity on central axis is suitable for characterizing the strength of oceanic turbulence in some distance of beam propagation path. Larger $d$ results in high sensitivity of oceanic turbulence effect at longer distance. But it also loses the capability at close range gradually.

\section{Conclusion}\label{sec5}

In this paper, we have investigated the evolution of average intensity of dual coherent Gaussian vortex beams with $+1$ and $-1$ topological charges respectively through weak oceanic turbulence. Considering the separation of two beams, the separation distance $2d$ enhances the oceanic turbulence effect on coherent superposition of light field. It has been shown that the maximum average intensity value is related to $d$, waist width $\sigma$, propagation distance $z$ and the strength of oceanic turbulence. When the form of initial light field is determined, the maximum intensity value of interference light field is more suitable for characterizing the strength of oceanic turbulence than single beam. However, the propagation distance $z$, $\sigma$, and non-monotonic relation to separation distance $d$ restrict it as a optimum parameter. Paying attention to the evolution of average intensity in x plane and y plane instead, we have found that the evolution of the intensity value on the central axis of two beams in y plane has monotonic relations to both propagation distance $z$ and strength of oceanic turbulence. Under the condition of weak oceanic turbulence, larger separation distance contributes to high sensitivity of oceanic turbulence effect at longer distance. But that leads to lower sensitivity at close range gradually. This feature is suitable for the non-contact measurement of oceanic turbulence strength by lateral scattering intensity, even for the turbulence strength varying with light propagation path.

\section*{References}
\bibliographystyle{iopart-num}
\bibliography{Jo}

\end{document}